\documentclass[aps,prd,onecolumn,groupedaddress,showpacs,nofootinbib,amssymb]{revtex4}
\usepackage[dvips]{graphicx}
\usepackage{amssymb}
\usepackage{amsmath}
\usepackage{graphicx}
\usepackage{amsfonts}
\usepackage{bm}
\begin{document}

\title{Inflationary scenario driven by type IV singularity in  $f(T)$ gravity}
\vspace{1cm}
\author{
H. F. Abadji $^{(a)}$\footnote{e-mail:jim2889@yahoo.fr},
M. G. Ganiou $^{(a)}$\footnote{e-mail:moussiliou\_ganiou@yahoo.fr},
M. J. S. Houndjo   $^{(a,b)}$\footnote{ e-mail: sthoundjo@yahoo.fr},
and J. Tossa $^{(a)}$\footnote{e-mail: joel.tossa@imsp-uac.org}
}
\vspace{1cm}
\affiliation{ $^a$ \, Institut de Math\'{e}matiques et de Sciences Physiques, 01 BP 613,  Porto-Novo, B\'{e}nin\\
$^{b}$\, Facult\'e des Sciences et Techniques de Natitingou, BP 72, Natitingou, B\'enin 
}
\begin{abstract}
In this paper, we investigate the effects  of  Type IV singularity through  $f(T)$ gravity description of inflationary universe, where $T$ denotes the torsion scalar. With the Friedmann equations of the theory, we reconstruct a  $f(T)$ model according  to a given Hubble rate susceptible  to describe the inflationary era near the type IV singularity. Moreover, we  calculate the Hubble flow parameters in order to determine the dynamical evolution of the cosmological system. The results show that some of the Hubble flow parameters are small near the Type IV singularity and become singular at  Type IV Singularity, indicating that a dynamical instability of the cosmological system occurs a that point.  This means that the dynamical cosmological evolution up to that point, ceases to be the final attractor since the system is abruptly interrupted. Furthermore, by considering the $f(T)$  trace anomaly equation and the slow-roll conditions,  we deal with the de Sitter inflationary description of the reconstructed model. As results, the model leads to a conditional instability, view as the source of the graceful exit from inflation. Our theoretical  $f(T)$  description based on slow-roll parameters not only confirms some observational data on spectral index and the scalar-to-tensor ratio from Planck data and BICEP$2$/Keck-Array data, but also shows the property of  $f(T)$ gravity in describing the  early and late-time evolution of our universe. 
\end{abstract}


\maketitle


\def\pp{{\, \mid \hskip -1.5mm =}}
\def\cL{\mathcal{L}}
\def\be{\begin{equation}}
\def\ee{\end{equation}}
\def\bea{\begin{eqnarray}}
\def\eea{\end{eqnarray}}
\def\tr{\mathrm{tr}\, }
\def\nn{\nonumber \\}
\def\e{\mathrm{e}}

\section{Introduction}
Unquestionably, the inflation era is of profound importance for the description of the primordial cosmological  evolution of our universe \cite{1}-\cite{14'} and many theoretical frameworks can successfully incorporate various version of this early-time acceleration.  With regards to the modified gravity description of inflation, it is possible to describe early and late-time acceleration within the same theoretical framework. The theoretical framework of $f(T)$ gravity has proved to be quite useful in cosmological and also astrophysical applications. Particularly, late-time acceleration in $f(T)$ gravity was studied in Ref \cite{2}-\cite{25'}. According to  the bouncing cosmology \cite{3}-\cite{32'}, the universe presents a contracted phase which ends up to a minimal radius, and after that starts an expanded phase. This means that the collapse feature of the universe never occurs, reveling the impossibility of initial singularity and also inflationary bouncing cosmology scenarios were studied in \cite{4}. Also various astrophysical aspects of $f(T)$ gravity were addressed in \cite{5,6} and in addition the thermodynamics of $f(T)$ and other modified gravities were studied in \cite{7}. In view of the various successful description of $f(T)$ gravity in both at local and global scales in the Universe, with this work we aim to investigate the implications of a Type IV singularity on the inflationary dynamics of vacuum $f(T)$ gravity. We attach attention to cosmological bounce including Type IV singularity finite time that can be generated from a general pure $f(T)$. We assume that the bouncing point is the same where the Type IV singularity occurs, and find out the algebraic form $f(T)$ function accordingly. In addition we shall calculate these corresponding Hubble flow(also called slow-roll) parameters \cite{8}, and we will investigate what are the effects of the Type singularity on the Hubble flow parameters. The Hubble flow govern the inflationary dynamics, so if the dynamics are affected directly by the presence of the Type IV singularity, this would be a clear indicator that the dynamics are is interrupted or modified at the singularity point, and therefore the final attractor solution can be changed at exactly the singularity point. As we shall demonstrate, the Hubble flow parameters are small for small cosmic times, but some of the them diverge at the singularity \cite{9}. we interpret the instability of the dynamical evolution of the cosmological system as an indication that inflation ends at the moment that the singularity occurs. The Type IV singularity acts as an indicator  of another possible mechanism for graceful exit from inflation in $f(T)$ gravity.\\
 
The present work is organized as  follow. In section \ref{sec1}, we make a brief summary of $f(T)$ gravity at the background level. In section \ref{sec2}, we present the time-like and finite-time singularities and conventions . The construction of a $f(T)$ model has been performed in \ref{mou} . In addition, in section \ref{sec3}, the stability of the constructed $f(T)$ solution near the bounce is realized. We describe the late-time cosmological evolution in  \ref{sec4} and the  conclusion ends the production  .


\section{ A brief summary $f(T)$ gravity }\label{sec1}
 The $f(T)$ gravity is a modified version of  Teleparallel theory of gravity that can be qualified as "torsion theory". Here,  we expose some definitions of torsion and associated quantities which will be useful to review $f(T)$ gravity before involving it in our present investigation.
In this formulation, the dynamical variable is the vierbein field $e_{i}(x^\mu); i=0,1,2,3$, which forms an orthonormal basis in the tangent space at each point $x^\mu$ of the manifold.  One has  $e_i.e_j=\eta_{ij}$ where $\eta_{ij} =diag(1,-1,-1,-1)$ is the Minkowsky metric. In a local coordinate basis $\partial_\mu $,  one can write $e_i=e_i\;^\mu\partial_\mu$ where $ e_i\;^\mu,\mu=,0,1,2,3$ represent the components of the vectors $e_i$. We use  the Latin indices  the tangent space and the Greek indices for the coordinates on the manifold (the space-time). The tangent space is Minkowskian and the relation between its metric and space-time metric is given by $g_{\mu\nu}=\eta_{ij}e^i\;_\mu(x)e^j\;_\nu(x)$, where $e^a\;_\mu e_a\;^\nu=\delta^\nu_\mu \;et\; e^a\;_\mu e_b\;^\mu=\delta^a_b\ $. 
In Teleparallel gravity, an important notion resulting from this theory is the condition of absolute parallelism which leads to the weitzenb\"{o}ck connection.  The curvature-less weitzenb\"{o}ck connection is considered  as the fundamental connection of the theory(contrarily to General Relativity (GR) which is based on the torsion-less Levi-Civita connection).  This Teleparallel connection is  given by 
\begin{equation}\label{h1}
\Gamma^\lambda\;_{\mu\nu}=e_a\;^\lambda\partial_\mu e^a\;_\nu
\end{equation}
The weitzenb\"{o}ck preserves the torsion tensor whose expression is given by 
\begin{equation} \label{h2}
T^\lambda\;_{\mu\nu}=\Gamma^\lambda\;_{\mu\nu}-\Gamma^\lambda\;_{\nu\mu}=e_a\;^\lambda(\partial_\mu e^a\;_\nu -\partial_\nu e^a\;_\mu) \neq0
\end{equation}
It is a non-vanishing and naturally antisymmetric  tensor. In GR, it is postulated that $T^\lambda\;_{\mu\nu}=0$. 
The action of teleparallel (TT) gravity is defined by 
\begin{equation}\label{h3}
	S=\frac{1}{2\kappa^2}\int d^{4}xeT,
\end{equation}
where $e=\det{e^a\;_\mu} =\sqrt{-g}$ and  $T$, the torsion scalar defined by
\begin{equation}
\label{h6}
T=S_\beta\;^{\mu\nu} T^\beta\;_{\mu\nu},
\end{equation} 
with 
\begin{equation}
S_\beta\;^{\mu\nu}=\frac{1}{2}\big (K^{\mu\nu}\;_\beta + \delta^\mu_\beta T^{\alpha\nu}\;_\alpha -\delta^\nu_\beta T^{\alpha\mu}\;_\alpha\big).
\end{equation}
 The new tensor representation $K^{\mu\nu}$ used in this previous expression is call the contorsion tensor and gives the difference between the Weitzenb$\ddot{o}$ck connection and the Levi-Civita connection.  It is expressed as \cite{kesti}
\begin{equation}
K^{\mu\nu}\;_\beta=-\frac{1}{2}\big (T^{\mu\nu}\,_\beta-T^{\nu\mu}\,_\beta-T_\beta\;^{\mu\nu}\big).
\end{equation}
Although this geometric difference between Teleparallel and GR, it is shown that their Lagrangian densities $T$ and $R$  are related by \cite{12}
\begin{equation}
R=-(T+2\nabla^{\mu}T^{\rho}\;_{\mu \rho}),
\end{equation} 
where $\nabla^{\mu}$ is the covariant derivative such that the total divergence $\nabla^{\mu}T^{\rho}\;_{\mu \rho}$ does not have any effect on the motion equations in the two theories. Accordingly, Teleparallel and GR are equivalent theories. 
In fact,  the idea of $f(T)$ gravity is to generalize $T$ by an arbitrary function $f(T)$, which is similar in spirit to the generalization of the Ricci scalar $R$ in the Einstein-Hilbert action to a function $f(R)$ \cite{Sergy}.

\begin{equation}\label{h4}
		S=\frac{1}{2\kappa^2}\int d^{4}x e(f(T)) + S_m
\end{equation}
where $S_m$ is the action of matter.  The variation of the action (\ref{h4}) with respect to the tetrad $e^a\;_\nu$ leads to the following field equation:
\begin{equation} \label{h5}
	S_{\mu}\;^{\rho\nu}\partial_\rho(T)f_{TT}+[e^{-1}e^{i}\;_\mu\partial_\rho(ee_i\;^\alpha S_\alpha\;^{\rho\nu)}-T^\alpha\;_\lambda\mu S_\alpha\;^{\nu\lambda}]f_T -\frac{1}{4}\delta_\mu^\nu f=-4\pi\mathcal{T}_\mu^\nu
\end{equation}
where $f_T$ and $f_{TT}$ denote the first and the second order derivatives of $f(T)$ with respect to the torsion scalar $T$, and $\mathcal{T}^{\nu}_{\mu}$ is the energy-momentum tensor constructed by the matter field Lagrangian.\\

If we assume the background manifold to be a spatially flat Friedmann-Robertson-Walker($FRW$) universe then the vierbein ou the tetrad fields takes the form
\begin{equation} 
\label{h7} 
e^a_{\:\:\:\mu}=diag(1,a,a,a);
\end{equation}
with $a$ the scale factor of the universe. Hence the dual vierbein $e_a\;^\mu=diag(1,\frac{1}{a},\frac{1}{a},\frac{1}{a})$ and the determinant $e=a^3$. From the relation between the tetrad and the metrics, one can immediately see that this choice gives rise to the well-known $FRW$ metric
\begin{equation} \label{h8}
	ds^2=dt^2 - a^2(t)\left(dx^2+dy^2+dz^2\right). 
\end{equation}
Moreover, assuming the matter as a perfect fluid, its energy momentum tensor takes the form
\begin{equation} 
\mathcal{T}_{\mu\nu}=(\rho + p)u_\mu u_\nu-p g_{\mu\nu},
\end{equation}
where $p$,\quad $\rho$, and $u^\mu$ are the pressure, energy density and four velocity of the matter fluid.  The scalar torsion  $T$  reads
\begin{equation} \label{h13}
	T=-6H^2.
\end{equation}
We have introduced the Hubble parameter $H\equiv\frac{\dot{a}}{a}$ to describe the expansion rate of the universe, where $\dot{a}$ denotes a derivative with respect  to the cosmic time.\par
 According to the metric (\ref{h8}), the field equation (\ref{h5}) gives two independant Friedmann  modified equations: 
\begin{eqnarray} \label{h9}
	-Tf_{T}(T)+\frac{1}{2}f(T)&=&\kappa^2 \rho,\\
	2\dot{T}Hf_{TT}(T) + 2(\dot{H}+3H^2)f_{T}(T)+\frac{1}{2}f(T)&=&-\kappa^2p_,
\end{eqnarray}
Assuming that the matter component is conserved, one has
\begin{equation} \label{h11}
\dot{\rho} +3H(\rho +p)=0,
\end{equation}
and obeys the barotropic equation of state (EoS)
\begin{equation}
	p=\omega\rho
\end{equation}
One can write Eq.(\ref{h9}) as 
functions of the effective energy density $\rho_{eff}$, and effective pressure $	p_{eff}$ as 
\begin{eqnarray}
\label{mousse1}
\rho_{eff}&=&\frac{3}{\kappa^2}H^2=\rho  + \rho_{T},\\
\label{mousse2}
 p_{eff}&=&-\frac{1}{\kappa^2}(2\dot{H}+ 3H^2)= p+ p_{T},
\end{eqnarray}
where 
\begin{eqnarray}
\label{h23}
\rho_T&=&\frac{1}{2\kappa^2} \big[ 2Tf_T -T- f(T) \big],\\
p_T&=&-\frac{1}{2\kappa^2} \Big[ 4\dot{H}(-2Tf_{TT}-f_{T}+1)]-\rho_T,\\
\end{eqnarray}

\section{Future finite-time singularities and conventions }\label{sec2}
In  this section, we recall some information  regarding finite time singularities. The finite time classification was done by using three physical quantities, the effective energy density $\rho_{eff}$, the effective pressure $p_{eff}$, the scale factor $a(t)$, and also the Hubble rate and it's higher derivatives in the following way \cite{10} 
\\
$\bullet$ Type I ( known as Big Rip): It is the most severe among finite time cosmological singularities from the phenomenological point of view. It occurs when the cosmic time approaches specific time ($t_s$), the scale factor $a$, the effective density $\rho_{eff}$ and the effective pressure $p_{eff}$ diverge, namely is : $a\rightarrow \infty$, $\rho{eff}\rightarrow\infty$, and $\arrowvert p_{eff}\arrowvert \rightarrow \infty$.\\
$\bullet$ Type II (known as Sudden Singularity): this singularity occurs when, as the cosmic time $t$ approaches $ts$, the scale factor and the effective energy density take bounded values, while the effective pressure diverges : $a\rightarrow a_s$, $\rho_{eff}\rightarrow\rho_s$ and $\arrowvert p_{eff}\arrowvert \rightarrow \infty$.\\
$\bullet$ Type III : this singularity occurs when, as the cosmic time $t$ approaches $ts$, only the scale factor remains finite, but both the effective energy density and the effective pressure diverge:$a\rightarrow a_s$, $\rho_{eff}\rightarrow\infty$ and $\arrowvert p_{eff}\arrowvert \rightarrow \infty$.\\
$\bullet$ Type IV : the most mild from a phenomenological point of view among all the three aforementioned types finite time of singularities, since in this case, the Universe can smoothly pass through it, with all the physical quantities that can be defined on the three dimensional spacelike hyper-surface  $t=ts$ remaining finite. Particularly: $a\rightarrow a_s$, $\rho_{eff}\rightarrow\rho_s$ and $\arrowvert p_{eff}\arrowvert \rightarrow p_s$. Moreover the Hubble rate and its first derivative also remain finite, but the higher derivatives, or some of these lasts diverge. \\ 
Our present investigation will be based on  the Type IV singularity. At that point, the bouncing cosmology and its evolution are studied in context of $f(T)$ gravity.
\section{Reconstruction of $f(T)$ near the type IV singularity}\label{mou}

In the present section, we develop the general reconstruction scheme for modified gravity with $f(T)$ action. It is shown how any cosmology may define the implicit form of the function $f$. Starting from action (\ref{h4}), we first consider the proper Hubble rate $H$ which describes the inflationary Universe and we investigate which vacuum $f(T)$ gravity can generate such a cosmological evolution, emphasizing to cosmic times near the Type IV singularity. The reconstruction is not explicit, and it is necessary to solve the differential equation and algebraic function. It shows, however, that at least, in principle, we could obtain any cosmology by properly reconstructing a function $f(T)$ at theoretical level. The equivalent form of gravitationnal action (\ref{h4}) above in a vacuum Jordan frame $f(T)$ gravity is: 
\begin{equation} \label{h14}
	S=\frac{1}{2\kappa^2}\int d^4xef(T)
\end{equation}
After varying this above action with respect to the tetrad, and making using of $FRM$ metric, the trace of the corresponding Friedmann equations reads:
\begin{equation}
	-36H^2\dot{H}f_{TT}+3(4H^2+\dot{H})f_T  + f =0.
\end{equation}
    The reconstruction method followed here is based on  an auxiliary scalar field $\phi$ such that  the action (\ref{h14}) become:
    \begin{equation} \label{h15}
    	S=\int d^4xe[P(\phi)T + Q(\phi)].
    \end{equation}
    where $P$ and $Q$ are  function of the scalar field$\phi$. This reconstruction method consists here on finding the analytic dependence of $P(\phi)$ and $Q(\phi)$ on the torsion scalar. Then we assume that the scalar field does not have a kinetic term, we may take $\phi$ as an auxiliary field such that the variation of (\ref{h15}) with respect to this latter, gives
    \begin{equation} \label{h16}
    	P'(\phi)T + Q'(\phi)=0.
    \end{equation}
Here $P'(\phi)$ and $Q'(\phi)$ denote the derivatives of  $P$ and $Q$ with respect to $\phi$. The reconstructed algebraic function is searched in the following form:
\begin{equation} \label{h17}
	f(\phi(T))= P(\phi(T))T+ Q(\phi(T)).
\end{equation} 
By substituting (\ref{h17}) in .(\ref{h9}), one gets:
\begin{equation} \label{h18}
6H^2P(\phi) + Q(\phi)=0.
\end{equation} 
\begin{equation}\label{h19}
	2H\dot{P}+2\dot{H}P+6H^2P(\phi) + Q(\phi)=0.
\end{equation}
By eliminating $Q(\phi)$ from (\ref{h18})-(\ref{h19}), one obtains the following first order differential equation
\begin{equation}
		2H\dot{P}+2\dot{H}P=0.
\end{equation}
Taking  the auxiliary scalar field $\phi$ as the cosmic time $t$,  the resolution of above equation gives
\begin{equation} \label{h20}
	P(t)=\frac{c_{1}}{H(t)},
\end{equation}
with $c_{1}$ an arbitrary integration constant and $H(t)$, the Hubble rate which described the inflationary universe. According to  (\ref{h13}), we can write $P$ in the following form
\begin{equation}
	P(T)=c_1\sqrt{6}(-T)^{\frac{-1}{2}}
\end{equation}
Then   $P(T)$ in (\ref{h18}) gives rise to
\begin{equation}
	Q(T)=-c_{1}\sqrt{6}(-T)^\frac{1}{2}
\end{equation}
Hence, we easily reconstructed $f(T)$ of /(\ref{h17}) as:
\begin{equation} \label{h21}
	f(T)=-2c_1\sqrt{6}(-T)^\frac{1}{2}
\end{equation}
It is quite remarkable that this form of $f(T)$ obtained has been done without any specific form of Hubble rate and therefore may be used to study all the four types of singularity.\\
To solve the cosmological problem we want inflation to last for a sufficiently long time, so we shall be interested in the behavior of the solution near the Type IV singularity, that means that we will expand the reconstructed $f(T)$ gravity around the type IV singularity. The chosen Hubble rate which can describe the inflationary Universe is:
\begin{equation}\label{h22}
	H(t)= c_0+ f_0(t-t_s)^\alpha
\end{equation}
with the assumption that $c_0\gg f_0$ and also for the cosmic times near the inflationary era , it holds true that $ c_0\gg f_0(t-t_s)^\alpha$, for $\alpha > 0$. So in effect, near the Type time instance $t\simeq t_s$, the tensor scalar reads $T_{s}=-6c_0^2$ and the cosmological evolution is nearly the de Sitter. Also, the Type IV singularity occurs at $t=t_s$, as it can be seen from (\ref{h22}) and $H(T_s)=c_0$. Particularly, the cosmological evolution fallen from the relation(\ref{h22}), is determined from the values of the parameter $\alpha$, and for various values of $\alpha $, one obtains what follow,\\
$\bullet$ $\alpha <-1$ corresponds to Type I singularity.\\
$\bullet$ $-1<\alpha<0$ corresponds to Type III singularity\\
$\bullet$ $0<\alpha<1$ corresponds to Type II singularity\\
$\bullet$ $\alpha>1$ corresponds to Type IV singularity\\
Then the expanding solution around the Type IV singularity reads:
\begin{equation}
	f(T)=f(T_s)+(T-T_s)f'(T_s)+\frac{1}{2}(T-T_s)^2f''(T_s)+...,
\end{equation}
which yields,
\begin{equation}\label{h23}
	f(T)\simeq T + \frac{1}{36c_0^2}T^2-3c_0^2,
\end{equation} 
with $ f(T_s)=-12c_1c_0,\quad f'(T_s)=\frac{c_1}{c_0}\quad and \quad f''(T_s)=\frac{c_1}{12c_0^3}$. We have also made  $c_1=\frac{2}{3}c_0$.\\
Moreover, the coefficient $\alpha=\frac{1}{36c_0^2}$ of $T^2$ higher order gravity of (\ref{h23}) contributes to the inflationary phase in the ordinary Starobinsky  inflation of the form $T+ \alpha T^2 -\Lambda$.
The direct comparison of the relation (\ref{h23}) with the de Sitter solution   \cite{44} relates the coefficient $\alpha$ to the cosmological constant $\Lambda$ by $\alpha=\frac{1}{12\Lambda}$ where $\Lambda=3c_0^2$

\section{  Stability analysis of the  reconstructed $f(T)$ Model} \label{sec3}
\subsection{Inflationary description by slow-roll parameters}
We investigate in this section   the inflationary universe description by defining the first slow-roll index \cite{11} and its running
\begin{equation} \label{h24}
	\epsilon_1=-\frac{d \ln H/dt}{H},\quad \epsilon_{N+1}=\frac{d \ln \epsilon_N /dt}{H}; \quad N \in Z
\end{equation}
Generally, the Hubble flow parameters of (\ref{h24}) constitute good indicators of the dynamical evolution because they practically  control the dynamics. Particularly, if $\epsilon_i \ll 1$, inflation occurs, and it stops when these become of order $\epsilon_i\sim 1$ \cite{resul11}-\cite{resul1}. According to the relations presented in (\ref{h24}), these parameters can be expressed as   

\begin{eqnarray}
	\epsilon_1&=&-\frac{\dot{H}}{H^2}\\
	 \epsilon_2&=& \frac{\ddot{H}}{\dot{H}H}-2\frac{\dot{H}}{H^2}\\ 
	\epsilon_3&=& \frac{\frac{4\dot{H}^2}{H^3}-\frac{3\ddot{H}}{H^2}-\frac{\ddot{H}^2}{H\dot{H}^2}+\frac{\dddot{H}}{H\dot{H}}}{H(-2\frac{\dot{H}}{H^2}+\frac{\ddot{H}}{H\dot{H}})} \\
	 \epsilon_4&=&-\frac{4\dot{H}^4+3H\dot{H}^2\ddot{H}-H^2\ddot{H}^2-H^2\dot{H}\dddot{H}}{2H^2\dot{H}^3-
	 	H^3\dot{H}\ddot{H}}+ \frac{13\dot{H}^3\ddot{H}-H\dot{H}^2\dddot{H}-H^2\ddot{H}\dddot{H}+H\dot{H}(-8\ddot{H}^2+H\ddddot{H})}{4H\dot{H}^4-
	 	3H^2\dot{H}^2\ddot{H}-H^3\ddot{H}^2+H^3\dot{H}\dddot{H}}\nonumber\\
\end{eqnarray}  
  By making using the  Hubble rate of (\ref{h22}) whose parameters are  constrained  by $f(T)$ gravity of Eq.(\ref{h23}), emphasizing to their form near the Type IV singularity at $t-t_s$, these parameters reads:
\begin{eqnarray}
	\epsilon_1&=&-\frac{f_0(t-t_s)^{-1+\alpha}\alpha}{(c_0+f_0(t-t_s)^\alpha)^2},\\
\epsilon_2&=&\frac{\alpha(-1+\alpha)f_0(t-t_s)^{-2+\alpha}}{\alpha f_0(t-t_s)^{-1+\alpha}(c_0 +f_0(t-t_s)^\alpha)}-\frac{2\alpha f_0(t-t_s)^{-1+\alpha}}{(c_0+f_0(t-t_s)^{\alpha})^2},\\
	\epsilon_3&=&\frac{-c_0^2(-1+\alpha)-3\alpha(-1+\alpha) c_0f_0(t-t_s)^\alpha +4\alpha^2 f_0^2(t-t_s)^{2\alpha}}{c_0^2(t-t_s)[c_0(-1+\alpha)-2\alpha f_0(t-t_s)^\alpha]},\\
	\epsilon_4&=&\frac{A(t)}{B(t)},
	\end{eqnarray}
with
\begin{eqnarray}
A(t)&=&-c_0^4(-1+\alpha)^2 -12c_0^2f_0^2(t-t_s)^{2\alpha} (-1+\alpha)^2 \alpha^2+26c_0 f_0^3(t-t_s)^{3\alpha} (-1+\alpha)\alpha^3\nonumber\\ 
&&-16f_0^4(t-t_s)^{4\alpha} \alpha^4+2c_0^3f_0(t-t_s)^\alpha \alpha(-2+3\alpha-2\alpha^2+\alpha^3)\nonumber\\
B(t)&=& c_0^2(t-t_s)(c_0(-1+\alpha)-2f_0(t-t_s)^\alpha \alpha)(c_0^2(-1+\alpha)+3c_0 f_0(t-t_s)^\alpha(-1+\alpha)\alpha\nonumber\\
&&-4f_0^2(t-t_s)^2\alpha \alpha^2)\nonumber
\end{eqnarray}	

So near the type IV singularity they become:
\begin{equation} \label{h25}
	\epsilon_1\simeq -\frac{f_0(t-t_s)^{-1+\alpha}\alpha}{c_0^2}; \quad \epsilon_2 \simeq \frac{-1+\alpha}{t-t_s} ;\quad \epsilon_3 \simeq  \frac{-1}{c_0(t-t_s)} ;\quad \epsilon_4 \simeq \frac{-1}{c_0(t-t_s)}
\end{equation}
Since $\alpha>1$ and $\frac{f_0}{c_0}\ll 1$, the parameters $\epsilon_1$ satisfies $\epsilon_1 \ll 1$ while at the Type IV singularity it becomes equal to zero.
As  we can see from (\ref{h25}), for times near the singularity $t<t_s$ the parameters $\epsilon_2, \quad \epsilon_3,\quad \epsilon_4$ satisfy $\arrowvert \epsilon_i \arrowvert \ll 1$, while at the Type IV singularity, that is $t=t_s$ they diverge due to the presence of the terms $\sim (t-t_s)$ which is singular at $t=t_s$ and this, regardless the value of $\alpha$\\
We are confronted to the following physical picture: the Hubble parameters that control the dynamics of the inflationary solution for times $t<t_s$, take very small values $\arrowvert \epsilon_i \arrowvert \ll 1$, but at the Type IV singularity three of these diverge regardless the value of $\alpha$. The infinite singularity of parameters $\epsilon_3, \epsilon_4$ clearly indicates a dynamical instability of the cosmological system. Moreover, we could say that the dynamical evolution is interrupted violently, so this could mean that the inflationary solution that described the dynamical cosmological evolution up to that point, cease to be the final attractor of the dynamical system. therefore, we could loosely state that this singularity could be an indicator that inflation ends and indicates the graceful exit from inflation.      
\subsection{ The de Sitter solution and instability from the trace anomaly equation}
We investigate here the  de sitter inflationary scenario description driven by the  reconstructed $f(T)$ model in (\ref{h23}). We emphasize here that by using the  2nd FRW equation of the system (\ref{h9}) in absence of matter, this $f(T)$ model leads to the de Sitter solution $H(t)=c_0=$constant. This approach is the currently used approach followed by \cite{9}. After recovering the de Sitter solution, the second step of this approach consists  to perturb the  2nd FRW equation around the de Sitter solution in order to study the instability of the solution view as the source of the graceful exit from inflation. Unfortunately, the approach fails at this level because the perturbative equation doesn't lead to physical solution. Consequently, we are going to make  resort to the trace anomaly equation in the same approach as  \cite{42}.  We examine the influence of the trace anomaly on our    inflationary description via the constructed quadratic model (\ref{h23}).
The trace $T_T$  of the energy-momentum tensor in the torsion scalar can be obtained from the equations (\ref{mousse1}) and (\ref{mousse2}) by
\begin{equation}
	T_T=-\rho_T+3p_T=-\frac{2}{\kappa^2}(6H^2 + 3\dot{H}-f-3\dot{H}f_T-12H^2f_T+36H^2\dot{H}f_{TT}).
\end{equation}
By taking into account the contribution from trace anomaly one gets \cite{42}: 
\begin{equation}\label{h30}
	-\frac{2}{\kappa^2}(-f-3\dot{H}f_T-12H^2f_T+36H^2\dot{H}f_{TT})-\Big(\frac{2}{3}\tilde{b}+\tilde{b}''\Big) \Big(\frac{d^2}{dt^2}+3H\frac{d}{dt}\Big)(12H^2+6\dot{H})+24\tilde{b}'\Big(H^4+H^2\dot{H}\Big)=0.
\end{equation}
with $\tilde{b}, \tilde{b}' and\;\tilde{b}''$ constants and the prime is not the derivative with respect to $T$ but just a superscription as a notation. For  $N$ scalars, $N_{1/2}$ spinors, $N_1$ the vectors fields, $N_2$ (=0,1) gravitons and $N_{HD}$ higher-derivative conformal scalars :\\
\begin{equation} \label{h27}
	\tilde{b}=\frac{N+6N_{1/2}+12N_1+611N_2-8N_{HD}}{120(4\pi)^2},\;\tilde{b}'=-\frac{N+11N_{1/2}+62N_1+1411N_2-28N_{ND}}{360(4\pi)^2}.
\end{equation}
we should note that $\tilde{b}>0$ and $\tilde{b}'<0$ 
for the usual matter, except higher-derivative conformal scalars and $\tilde{b}''$ can be shifted by the finite re-normalization and being an arbitrary coefficient. In the de Sitter space, $ H=H_{dS}=Constant $ and the (\ref{h27}) reduces to an algebraic form
\begin{equation}\label{h28}
	-\frac{2}{\kappa^2}(-f-12H^2f_T) +24\tilde{b}'H^4_c=0.
\end{equation}
 By putting in (\ref{h28}) the reconstructed quadratic model $f(T)=T+\alpha T^2 -\Lambda$ of (\ref{h23})  one has 
\begin{equation}
	-6H^2_{dS}+\alpha (-6H^2_{dS})^2- \Lambda + 12H^2_{dS} \Big[1+2\alpha(-6H^2_{dS}) \Big]+12\tilde{b}'H^4_{dS}\kappa^2=0,
\end{equation}
where $\alpha=\frac{1}{12\Lambda}>0$,\, $\Lambda=3c_0^2$. It follows
 \begin{equation}
 	H^2_{dS}=0 \qquad\text{and/or} \qquad \frac{9\alpha}{\kappa^2}-\tilde{b}'=\frac{1}{4H_{dS}^2\kappa^2},
 \end{equation}
 which imposes
\begin{equation} \label{h29}
H^2_{dS}=0 \qquad \text{and/or} \qquad \frac{9\alpha}{\kappa^2}-\tilde{b}'>0.
\end{equation}
Since $\alpha > 0$, the condition (\ref{h29}) is satisfied because $\tilde{b}'< 0$ so the de Sitter inflation can be realized and the corresponding de Sitter solution is $H_{dS}=\sqrt{\frac{1}{36 \alpha-4\kappa^2\tilde{b}'}}$.
Now, we aim to study the instability of the de Sitter solution source of  the end of inflation. To do so, we introduce the reconstructed model in the gravitational field (\ref{h30}) including the trace anomaly
\begin{eqnarray}\label{h31}
	&&\frac{6}{\kappa^2}\Big[2H^2-36\frac{1}{(36c_0^2)}H^4-c_0^2+\dot{H}-36\frac{1}{(36c_0^2)}H^2\dot{H}- \kappa^2 \Big(\frac{2}{3}\tilde{b}'+\tilde{b}''\Big)\Big(\dddot{H}+7H\ddot{H}+12H^2\dot{H}+4\dot{H}^2\Big)\nonumber \\
	&&	+4\tilde{b}'\kappa^2\Big(H^4+H^2\dot{H}\Big)\Big]=0.
\end{eqnarray}
We remark that this resulting equation of the reconstructed model generalizes one obtained in \cite{42}.To analyze inflation occurring in the reconstructed model, we study the stability of the de Sitter solution towards linear perturbations of the following form:
\begin{equation} \label{h34}
	H(t)=H_{dS}(1-\delta(t))\qquad \delta\ll 1
\end{equation}
We have to calculate the amplitude of the perturbations. Its sign is crucial since the stability or not of the Hubble parameter depend on it.  $\delta=0$ corresponds logically to the  inflation era or its beginning. We consider here only the unstable case $\delta>0$ i.e $H<H_{dS}$. So,from (\ref{h31}), we find:
\begin{eqnarray}\label{h32}
&&0=\kappa^2H_{dS}\Big(\frac{2}{3}\tilde{b}'+\tilde{b}''\Big)\dddot{\delta}(t)+7\kappa^2H_{dS}^2\Big(\frac{2}{3}\tilde{b}'+\tilde{b}''\Big)\ddot{\delta}(t)-\Big[1-36\Big(\frac{1}{36c_0^2}\Big)H_{dS}^2-12\kappa^2H_{dS}^2\Big(\frac{2}{3}\tilde{b}'+\tilde{b}''\Big) \nonumber\\
&&+4\tilde{b}'\kappa^2H^2_{dS}\Big]H_{dS}\dot{\delta}(t)+4\tilde{b}'\kappa^2H^2_{dS}\Big]H_{dS}\dot{\delta}(t)+\Big[1-36\Big(\frac{1}{36c_0^2}\Big)H_{dS}^2+4\tilde{b}'\kappa^2H^2_{dS}\Big]4H^2_{dS}\delta(t)
\end{eqnarray}
Expressing $\delta(t)$ as $\delta(t)=\exp(\sigma t)$, the de Sitter solution is unstable only if the eigenvalue $\sigma$ is positive leading to the growing of $\delta$ with the cosmic time. Substituting $\delta(t)=\exp(\sigma t)$ in (\ref{h32}), one has
\begin{eqnarray}
	&&\sigma^3 + 7H_{dS}\sigma^2-\frac{1-36\Big(\frac{1}{36c_0^2}\Big)H_{dS}^2-12\kappa^2H_{dS}^2\Big(\frac{2}{3}\tilde{b}'+\tilde{b}''\Big)+4\tilde{b}'\kappa^2H^2_{dS}}{	\kappa^2\Big(\frac{2}{3}\tilde{b}'+\tilde{b}''\Big)}\sigma\nonumber \\
	&&+ \frac{\Big[1-36\Big(\frac{1}{36c_0^2}\Big)H_{dS}^2+4\tilde{b}'\kappa^2H^2_{dS}\Big]4H_{dS}}{	\kappa^2\Big(\frac{2}{3}\tilde{b}'+\tilde{b}''\Big)}=0
\end{eqnarray}
We assume that $\dddot{H}\ll H\dot{H}$ since $H$ variation is slow during inflation. We resolution  of the second order equation shows:
\begin{equation} \label{h33}
	\sigma=\frac{1}{2}\Big[\Big(\frac{7Q-12H_{dS}^2}{7H_{dS}}\Big)\pm\sqrt{D}\Big],
\end{equation}
with 
\begin{equation}
	Q=\frac{1-36(\frac{1}{36c_0^2})+4\tilde{b}'\kappa^2H^2_{dS}}{7[(\frac{2}{3}\tilde{b}+\tilde{b}'')]\kappa^2}
\end{equation}
\begin{equation}
	D=\Big(\frac{7Q-12H^2_{dS}}{7H_{dS}}\Big)^2-16Q
\end{equation}
So, due to the fact that $\tilde{b}>0$, $\tilde{b}'<0$, $\tilde{b}''$ an arbitrary constant, $Q$ can take a negative value. Then, when $Q<0$, $D$ becomes positive and we can choose  $\sigma>0$ with the $"+"$ sign in (\ref{h33}). With $\sigma>0$,the amplitude of the perturbation diverges in time and thus from (\ref{h34}) that $H$ decreases. This means that the de Sitter solution is unstable and its inflation can end.
\subsection{The Hubble slow-soll parameters for a type IV singular evolutions}
\begin{figure}[h] 
	\centering
		\includegraphics[width=10cm, height=10cm]{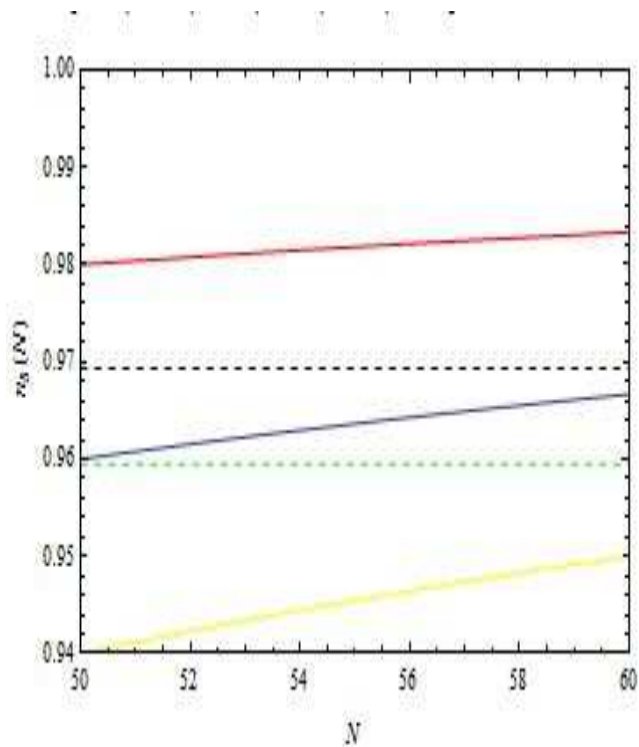}
	\caption{Spectral index $n_s$ for three values of the parameter $\alpha$: the red, the blue and the yellow curves  correspond to $\alpha=2$, $\alpha=3$ and  $\alpha=4$, respectively.  Furthermore, the limits of the value of $n_s$ , namely $n_s=0.9595$ and $n_s=0.9693$,  predicted by Planck data 2016 are represented by the dashed lines}
	\label{fig1}
\end{figure}

\begin{figure}[h]
	\centering
	\includegraphics[width=10cm, height=10cm]{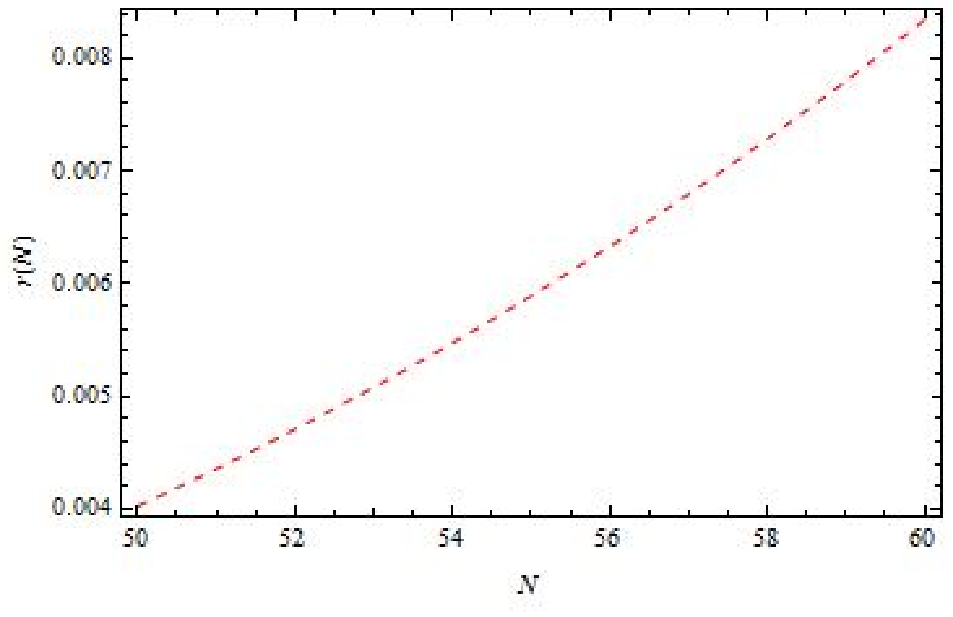}
	\caption{ The graph shows the scalar-to-tensor ratio near the type IV singularity for $\alpha=3$, $f_0=0.005$ and $c_0=8$}.
	\label{fig2}
\end{figure}

In the previous section, we used the Hubble flow parameters of  (\ref{h24}) to show that an instability  occurs in the case the cosmological evolution in $f(T)$ gravity. This instability was actually an infinite instability, since almost the Hubble  flow parameters contain singularities that occur at the Type IV singularity. In this section we discuss two Hubble slow-roll parameters $\epsilon_H$ and $\eta_H$ \cite{Liddle} in order to investigate the inflationary dynamics evolution  for completeness.
The concerned Hubble slow-roll parameters are expressed by
\begin{equation}
	\epsilon_H=-\frac{\dot{H}}{H^2}, \quad \eta_H=-\frac{\ddot{H}}{2H\dot{H}},
\end{equation}
with the Hubble rate (\ref{h22}), they become
\begin{equation}
	\epsilon_H=-\frac{f_0(t-t_s)^{-1+\alpha}\alpha}{[c_0+f_0(t-t_s)^\alpha]^2},\quad \eta_H=-\frac{-1+\alpha}{2[c_0+f_0(t-t_s)^\alpha](t-t_s)}
\end{equation}
and near the Type IV singularity, they give
\begin{equation}
	\epsilon_H=-\frac{f_0(t-t_s)^{-1+\alpha}\alpha}{c_0^2},\quad \eta_H=-\frac{-1+\alpha}{2c_0(t-t_s)}.
\end{equation}
These relations provide that in the case of a Type IV singularity, with $\alpha>1$, the parameter $\epsilon_H$ is regular at the singularity whereas $\eta_H$ blows up at the Type Singularity. The corresponding observables namely
the spectral index and the  tensor-to-scalar ratio  of the primordial curvature perturbations are expressed by \cite{Liddle}
\begin{equation}\label{g1}
	n_s=1-4\epsilon_H-2\eta_H+2\epsilon_3-2\epsilon_4 ,\quad r=48 \epsilon_H^2. 
\end{equation}
Considering that  $t<t_s$, we define the e-folding number as
\begin{equation}
	N=\int_{t}^{ts} H(t')dt'= -\frac{f_0}{\alpha+1}(t-t_s)^{\alpha+1}-c_0(t-t_s),
\end{equation}
so near the Type IV singularity which occurs at $t_s$, one has
\begin{equation}
	N=-c_0(t-t_s)
\end{equation} 
and then the observables of (\ref{g1}) become  
\begin{equation}\label{film}
n_s=1+\frac{1-\alpha}{N} ,\qquad r=\frac{48\alpha^2(-\frac{N}{c_0})^2f_0^2}{c_0^2N^2}
\end{equation}
As it can be seen from  the relations in (\ref{film}), the spectral index, near the type IV singularity, depends only from the parameter $\alpha$ whereas the scalar-to-tensor ratio still keeps all the parameters of the inflationary model (\ref{h22}) . The recent set of values of the spectral index fixed by the Planck data 2016 \cite{planck} is    
    \begin{equation}
  n_s=0.9644\pm 0.0049. 
   \end{equation}
 We accordingly plot for different values of the parameter $\alpha$, its theoretical representation in (\ref{film}) via the figure (\ref{fig1}). It results that only  $\alpha=3$ (blue curve) gives the value of $ n_s$  within the allowed range 
   values set by Planck data. As example, for $N=60$, one has \\
   $ n_s=0.966\in [0.9693,0.9595]$.   The scalar-to-tensor  ratio (\ref{film})  near the type IV singularity is also plotted for $\alpha=3$ in the figure \ref{fig2}. The figure shows that the scalar-to-tensor is well within the observational constraints of the
   Planck collaboration \cite{planck} but also within the allowed values of the BICEP$2$/Keck-Array data \cite{bicep2}, which constrain   the scalar-to-tensor ratio as follows $r<0.10$ and $r<0.07$, respectively.  All these results describe well the inflation ($t<t_s$)  before the singularity in the approach as those followed by \cite{oikonomou}.

\section{Late-time cosmological evolution} \label{sec4}
The previous sections show that the model (\ref{h22}) can describe perfectly the inflationary dynamics of the universe up to its graceful exit at type IV singularity. Now, we are dealing with the  late-time behaviors of the cosmological model of (\ref{h22}). We recall here that the cosmological evolution described  by (\ref{h22}), has some appealing phenomenological features, since it can describe late-time and early time acceleration, and then it offers the possibility of a unified description of these acceleration eras \cite{9}. In order to confirm here this property in our $f(T)$ description, we  calculate the corresponding Equation of State (EoS) $\omega_{eff}$ and focus on its behavior near the Type IV singularity. By making using (\ref{mousse1}) and (\ref{mousse2}), one has
\begin{equation}
	\omega_{eff}= -1-\frac{2\dot{H}}{3H^2},
\end{equation}
which, by using  (\ref{h22}), becomes
\begin{equation}
	\omega_{eff}=-1-\frac{2f_0(t-t_s)^{-1+\alpha}\alpha}{3[c_0+f_0(t-t_s)^\alpha]^2}.
\end{equation}
At early times where   $c_0\gg f_0$, one has
\begin{equation}
	\omega_{eff}=-1-\frac{2f_0(t-t_s)^{-1+\alpha}\alpha}{3c_0^2}.
\end{equation}
So  the resulting EoS describes a nearly de sitter evolution ($	\omega_{eff}\sim-1$) when $t=t_s$, namely at the singularity. Furthermore at late time which means that the cosmic time is larger or equal to present times $ t_p\sim 10^{17}$sec, the EoS becomes
\begin{equation}\label{mousse3}
	\omega_{eff} \simeq -1-\frac{2t^{-1-\alpha}\alpha}{3f_0}.
\end{equation}
Since $t\geq t_p$,  the second term  of (\ref{mousse3}) is very small and so leads to $	\omega_{eff}\sim-1$. Consequently, the above equation of state describes a nearly de Sitter evolution. As conclusion, both early and late-time acceleration can be described by the same cosmological model (\ref{h22}).

\section{Conclusion}
We have considered in this work, the Type IV finite-time singularity in the Jordan frame, in the context of vacuum $f(T)$ gravity, where $T$ denoted the scalar tensor. Our inflationary description has been based on the reconstruction of the $f(T)$ model which described the dynamical evolution in the inflationary epoch near the Type IV singularity by considering  a specific form of Hubble rate. We demonstrated that the Hubble parameters or the slow-roll parameters  develop an infinite instability at the Type IV singularity, and this instability is an indicator that graceful exit from inflation is triggered. Furthermore, we have performed the de Sitter  stability of the corresponding  $f(T)$   model     near the Type IV singularity and validate our claim that graceful exit occurs near the Type IV singularity. In addition, our investigation shows that the de Sitter inflation occurs and finally the universe can exit from it because of the instability  coming from the trace anomaly. Our investigation lead to the values of the spectral index and the scalar-to-tensor ratio in perfect agreement with the range allowed by Planck data and BICEP2 data.\par 


{\bf Acknowledgments}: H. F. Abadji would like to thank DAAD for its financial support.

\end{document}